\documentclass[12pt]{article}
\usepackage{titlesec}
\usepackage{lineno}
\usepackage{lipsum}
\usepackage{changepage}
\usepackage{graphicx}
\usepackage{amsmath}
\usepackage{caption}
\usepackage{setspace}
\usepackage{color}
\usepackage{comment}
\usepackage{authblk}
\usepackage{bm}

\usepackage[numbers,sort&compress]{natbib}

\usepackage{times}

\usepackage[
 left=3cm,  
 right=2.6cm,  
 top=2.1cm,   
 bottom=2.1cm, 
]{geometry}

\definecolor{qired}{rgb}{0.6, 0, 0}

\DeclareCaptionFormat{boldnumberfirstsentence}{
 \textbf{#1#2}#3\par
}

\DeclareCaptionLabelSeparator{colonspace}{: }
\newcommand*\patchAmsMathEnvironmentForLineno[1]{%
 \expandafter\let\csname old#1\expandafter\endcsname\csname #1\endcsname
 \expandafter\let\csname oldend#1\expandafter\endcsname\csname end#1\endcsname
 \renewenvironment{#1}%
   {\linenomath\csname old#1\endcsname}%
   {\csname oldend#1\endcsname\endlinenomath}}%
\newcommand*\patchBothAmsMathEnvironmentsForLineno[1]{%
 \patchAmsMathEnvironmentForLineno{#1}%
 \patchAmsMathEnvironmentForLineno{#1*}}%
\AtBeginDocument{%
\patchBothAmsMathEnvironmentsForLineno{equation}%
\patchBothAmsMathEnvironmentsForLineno{align}%
\patchBothAmsMathEnvironmentsForLineno{flalign}%
\patchBothAmsMathEnvironmentsForLineno{alignat}%
\patchBothAmsMathEnvironmentsForLineno{gather}%
\patchBothAmsMathEnvironmentsForLineno{multline}%
}

\allowdisplaybreaks

\title{Evolutionary dynamics of memory-based strategies in repeated and structured social interactions}
\date{}

\author[1]{\fontsize{12}{14}\selectfont Ketian Sun}
\author[2,*]{\fontsize{12}{14}\selectfont Qi Su}
\author[1,3,*]{\fontsize{12}{14}\selectfont Long Wang}
\affil[1]{Center for Systems and Control, College of Engineering, Peking University, Beijing, 100871, China.}
\affil[2]{School of Automation and Intelligent Sensing, Shanghai Jiao Tong University, Shanghai, 200240, China}
\affil[3]{Center for Multi-Agent Research, Institute for Artificial Intelligence, Peking University, Beijing, 100871, China}
\affil[*]{Corresponding author. Email: \text{qisu@sjtu.edu.cn}; \text{longwang@pku.edu.cn}}

\begin{document}
\maketitle

\begin{abstract}
Human social life is shaped by repeated interactions, where past experiences guide future behavior. In evolutionary game theory, a key challenge is to identify strategies that harness such memory to succeed in repeated encounters. Decades of research have identified influential one-step memory strategies (such as Tit-for-Tat, Generous Tit-for-Tat, and Win-Stay Lose-Shift) that promote cooperation in iterated pairwise games. However, these strategies occupy only a small corner of the vast strategy space, and performance in isolated pairwise contests does not guarantee evolutionary success. The most effective strategies are those that can spread through a population and stabilize cooperation. We propose a general framework for repeated-interaction strategies that encompasses arbitrary memory lengths, diverse informational inputs (including both one’s own and the opponent’s past actions), and deterministic or stochastic decision rules. We analyze their evolutionary dynamics and derive general mathematical results for the emergence of cooperation in any network structure. We then introduce a unifying indicator that quantifies the contribution of repeated-interaction strategies to population-level cooperation. Applying this indicator, we show that long-memory strategies evolve to promote cooperation more effectively than short-memory strategies, challenging the traditional view that extended memory offers no advantage. This work expands the study of repeated interactions beyond one-step memory strategies to the full spectrum of memory capacities. It provides a plausible explanation for the high levels of cooperation observed in human societies, which traditional one-step memory models cannot account for.
\end{abstract}

\newpage
\section*{Introduction}
Cooperation in societies is often sustained by repeated interactions, which allow trust and reciprocity to accumulate over time \cite{melis2010human, rand2013human}. For example, repeated contractual relationships help organizations build and sustain mutual confidence \cite{vanneste2010repeated}, and ongoing interactions among lawyers promote cooperative behavior within legal communities \cite{johnston2002does}. In formal analysis, the prisoner’s dilemma has long served as the canonical model for studying the tension between individual incentives and collective welfare \cite{axelrod1981evolution, cook2005cooperation, efferson2024super, michel2024evolution, kawakatsu2024mechanistic, sun2025direct}. In one-shot encounters, rational self-interest drives defection, yielding the collectively worst outcome. By contrast, repeated interactions allow individuals to condition behavior on past outcomes, enabling reciprocity and sustained cooperation \cite{nowak2006five}. This principle underlies classic strategies such as Tit-for-Tat \cite{axelrod1984cooperation}, Generous Tit-for-Tat \cite{nowak1992tit}, and Win-Stay Lose-Shift \cite{nowak1993strategy}, which illustrate the potential of simple reciprocal rules to stabilize cooperation in iterated pairwise games.

From an evolutionary perspective, however, success in isolated dyadic encounters does not guarantee long-term prevalence of a strategy. What matters is whether a strategy can spread and sustain cooperation at the population level. Despite extensive research, most studies on repeated interactions are largely confined to narrow regions of the strategy space: some analyze only a few prominent strategies, whereas others focus exclusively on memory-$1$ forms, thereby exploring only a small fraction of the possibilities available within the repeated-interaction framework \cite{imhof2005evolutionary, wang2021evolution, stewart2013extortion, imhof2010stochastic, adami2013evolutionary, chen2023outlearning, akin2015you, hilbe2018evolution, park2022cooperation, tkadlec2023mutation}. In more general terms, strategies in iterated games can be described along three key features. The first is memory length $n$, which specifies the length of interaction history considered. The second is informational input, which distinguishes memory-$n$ strategies that track both players’ past actions from reactive-$n$ strategies that rely only on the opponent’s moves. The third is the decision rule, specifying whether responses are deterministic or stochastic. Although these features are central to strategy design, their relative importance for the evolution of cooperation is not yet well understood, leaving it unclear which aspects of memory-based strategies are most effective in sustaining cooperative behavior.
 
A further limitation of prior work is the assumption of well-mixed populations, where individuals interact at random and with equal probability \cite{hilbe2018evolution, park2022cooperation, tkadlec2023mutation}. In reality, human life is structured by social, spatial, or institutional ties \cite{taylor2007evolution, rand2014static, ohtsuki2006evolutionary, su2017spatial, sheng2024strategy, santos2005scale, li2020evolution, gokhale2010evolutionary, gracia2012heterogeneous}. Illustrative examples include online social networks, where ties represent friendships or follower relationships \cite{mislove2007measurement}, and academic collaboration networks, where links indicate co-authorship or shared projects \cite{newman2001structure}. Such network structures can exert a strong influence on the emergence and stability of cooperative behavior \cite{su2022evolutionmultilayer, su2018understanding, allen2019evolutionary, allen2017evolutionary, wang2024deterministic, sheng2023evolutionary, su2019evolutionary, wang2023high, su2019spatial}. A general framework for repeated-interaction strategies should therefore broaden the strategy set, while explicitly incorporating the effects of population structure.

In this work, we propose a general framework in which individuals play iterated prisoner’s dilemma games with their neighbors. Our approach is applicable to strategies of any memory length, employing diverse informational inputs and either deterministic or stochastic decision rules. 
Using this framework, we derive analytical conditions for the emergence of cooperation. We then introduce a unifying indicator that disentangles the effects of repeated-interaction strategies from network structure, enabling systematic comparisons across strategy families. Applying this indicator, we show that strategy evaluation should focus on evolving populations rather than isolated dyadic contests: while prior work suggested that extended memory offers no advantage in pairwise interactions, we find that long-memory strategies substantially enhance cooperation at the population level.
Taken together, these findings establish a general framework for understanding how repeated interactions shape cooperative outcomes, and provide theoretical insights into the mechanisms that sustain cooperation in human societies.

\section*{Model}
\begin{figure}[]
	\centering
	\includegraphics[width=\textwidth]{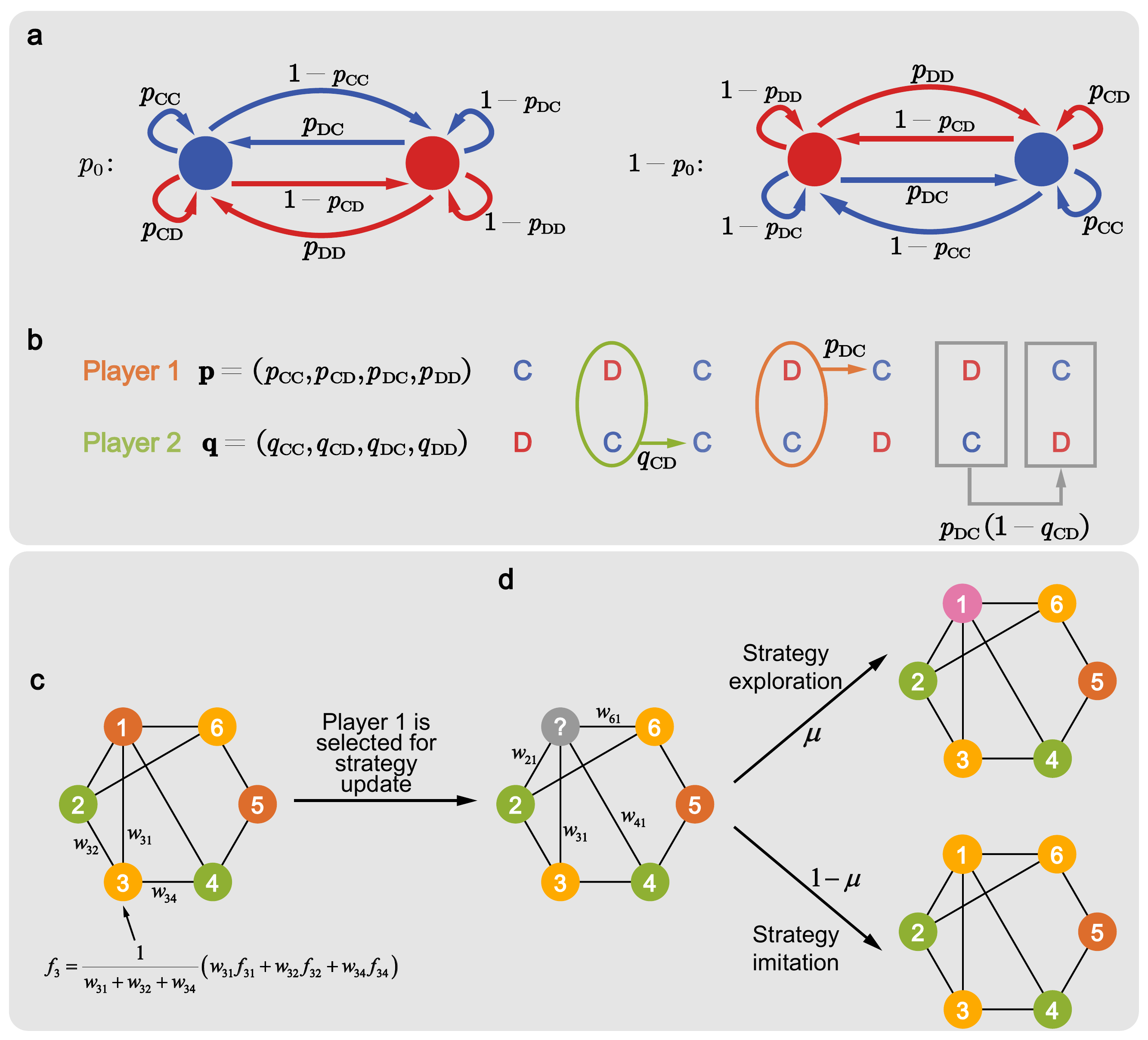}
	\caption{\textbf{Iterated games on networks.} \textbf{a}, Schematic of a memory-$1$ strategy. The initial cooperation probability is $p_0$, and $p_{a\tilde{a}}$ denotes the probability of cooperation given the previous actions of the player ($a$) and the opponent ($\tilde{a}$). Blue denotes cooperation and red denotes defection. Circles denote player actions, with the initial move on the left. Arrows indicate possible actions in the next round, with colors reflecting the opponent’s previous move and labels showing the corresponding probabilities. \textbf{b}, Example illustrating how past interactions influence behavior. Green and yellow arrows show the cooperation probabilities of players 1 and 2 following specific outcomes. The gray arrow marks a transition in the joint action profile from $\mathrm{D}\mathrm{C}$ to $\mathrm{C}\mathrm{D}$. \textbf{c}, Repeated interactions on a network. Nodes represent individuals, color-coded by strategy type. Iterated prisoner’s dilemma games are played along edges. The payoff of player 3 is an edge-weighted average, $f_3 = (w_{31}f_{31} + w_{32}f_{32} + w_{34}f_{34})/(w_{31}+w_{32}+w_{34})$. Fitness is given by $F_3 = \exp(\beta f_3)$. \textbf{d}, Strategy update. With probability $\mu$, player 1 explores by adopting a random strategy. Otherwise, player 1 imitates a neighbor, chosen with probability proportional to both edge weight and fitness. The probability of imitating player 3 is $w_{31}F_3 / (w_{21}F_2 + w_{31}F_3 + w_{41}F_4 + w_{61}F_6)$.} 
	\label{illus}
\end{figure}

We study a population of $N$ individuals structured as an undirected, weighted network. Each edge indicates a potential interaction, with weight $w_{ij}$ denoting the frequency of interaction between individuals $i$ and $j$. The normalized interaction frequency from $j$ to $i$ is $p_{ij} = w_{ij} / w_i$, where $w_i = \sum_k w_{ik}$ denotes the total edge weight of individual $i$.

Interactions occur along network edges, in the form of iterated prisoner’s dilemma games. In each round, players simultaneously choose to cooperate ($\mathrm{C}$) or defect ($\mathrm{D}$). A cooperator pays a cost $c$ to provide a benefit $b$ to the opponent, whereas a defector neither pays a cost nor provides a benefit. After each round, the game continues with probability $\delta \in (0,1]$, so the expected number of rounds is then $1/(1-\delta)$ (where $\delta=1$ corresponds to infinite repetition). This repetition enables strategies to condition present behavior on past outcomes.

Each individual adopts a strategy from a public set and applies the same decision rule to all neighbors. While our framework allows for arbitrary strategy sets, we analyze a family of classical strategies $\mathcal{S}^Y_{Xn}$, defined along three key features: (1) memory length $n$, which specifies the length of past interaction history considered; (2) informational input $X$, which distinguishes memory-$n$ ($Mn$) strategies that consider the actions of both players from reactive-$n$ ($Rn$) strategies that depend only on the opponent’s actions; and (3) decision rule $Y$, which differentiates deterministic ($\ast$, probabilities of either $0$ or $1$) from stochastic ($\dagger$, probabilistic) responses. For example, a memory-$1$ strategy is specified by an initial cooperation probability $p_0$ and four conditional probabilities $p_{a\tilde{a}}$, where $a$ and $\tilde{a}$ represent the previous actions of the player and the opponent, respectively (Fig.~\ref{illus}\textbf{a},\textbf{b}). For notational convenience, we write $\mathcal{S}^Y_X$ when $n=1$. To account for implementation errors, each intended action is assumed to be implemented incorrectly with probability $\varepsilon \in [0,0.5]$.

To model strategy evolution, we assume that updates occur after a sufficiently long period of repeated interactions. At that point, the frequencies of joint action profiles in each iterated game reach a stationary state. This allows payoffs to be computed. The payoff of individual $i$ is given by an edge-weighted average, $f_i = \sum_j p_{ij} f_{ij}$, where $f_{ij}$ denotes the stationary payoff from interactions with neighbor $j$. Fitness is defined as $F_i = \exp(\beta f_i)$, with $\beta \geq 0$ representing selection strength (Fig.~\ref{illus}\textbf{c}). At each update step, a randomly selected individual $j$ may revise their strategy: with probability $\mu$, player $j$ explores by adopting a random strategy from the set; otherwise, player $j$ imitates a neighbor $k$ with probability proportional to $w_{kj}F_k$, normalized over all neighbors $l$ of $j$, given by $w_{kj} F_k / \sum_l w_{lj} F_l$ (Fig.~\ref{illus}\textbf{d}).

\section*{Results}
\subsection*{Cooperation through repeated interactions on networks}

\begin{figure}[!t]
	\centering
	\includegraphics[width=\textwidth]{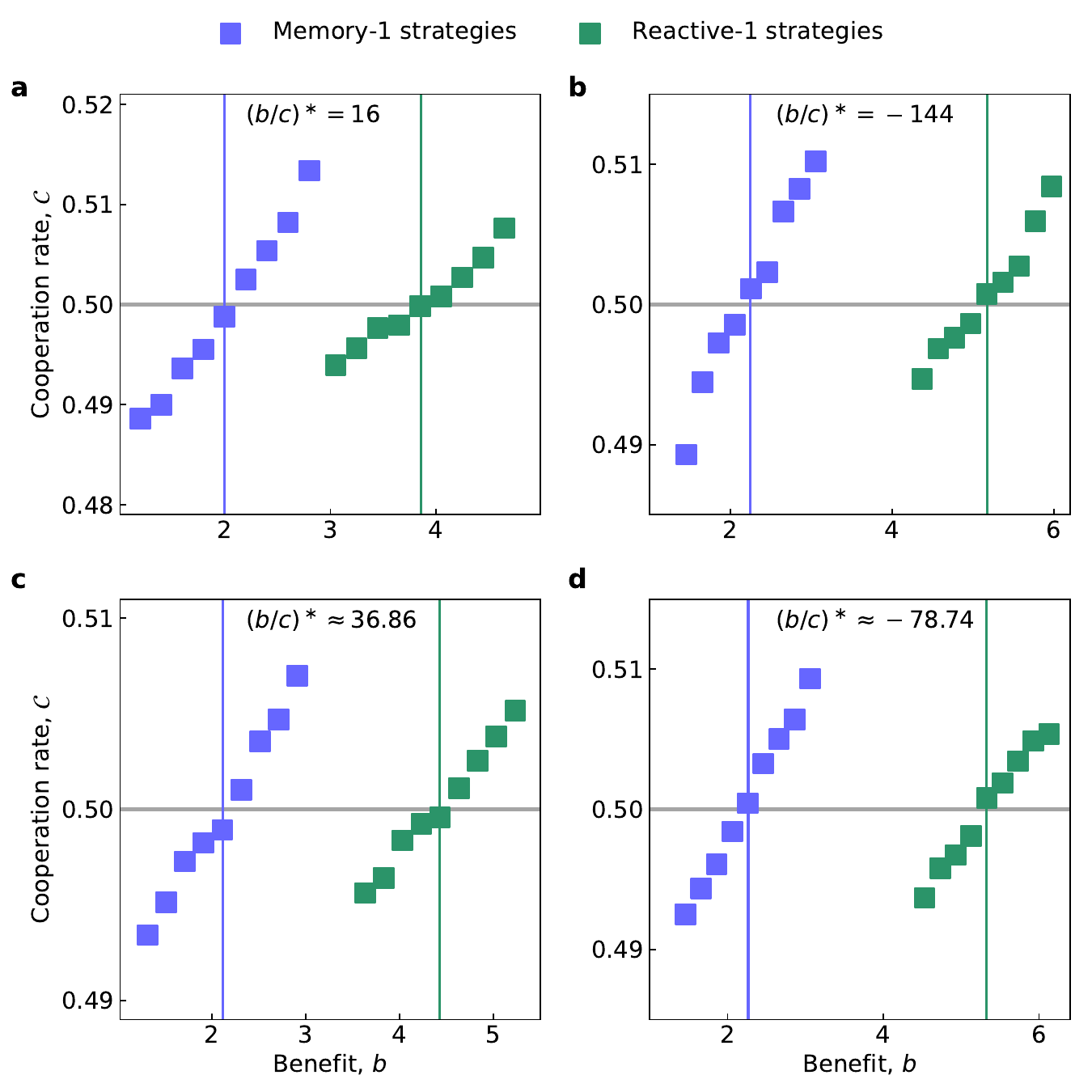}
	\caption{\textbf{Repeated interactions promote cooperation in structured populations.} \textbf{a}–\textbf{d}, Cooperation rates $\mathcal{C}$ plotted against the benefit $b$ on random regular (\textbf{a}, \textbf{b}) and scale-free (\textbf{c}, \textbf{d}) networks with average degree $k$. Cooperation is considered to emerge when $\mathcal{C} > 0.5$ (above the horizontal line). Symbols show simulation results, and the corresponding vertical lines indicate theoretical thresholds, for memory-$1$ (purple) and reactive-$1$ (green) strategy sets. The one-shot threshold $(b/c)^\ast$ is shown for reference. Parameters: $\delta=1$, $\varepsilon=0$, $N=50$, $k=10$ (\textbf{a}, \textbf{c}), $k=30$ (\textbf{b}, \textbf{d}), $c=1$, and $\beta=0.005$.} 
	\label{simul_det_inf_no}
\end{figure}

We define the population-level cooperation rate $\mathcal{C}^Y_{Xn}(\delta, \varepsilon)$ as the asymptotic frequency of cooperative behavior under strategy set $\mathcal{S}^Y_{Xn}$, continuation probability $\delta$, and error rate $\varepsilon$, evaluated in the limit of rare exploration, $\mu \rightarrow 0$ (see Methods). Cooperation is considered to emerge when $\mathcal{C}^Y_{Xn}(\delta, \varepsilon) > 0.5$. Under weak selection ($\beta \rightarrow 0$), the critical benefit-to-cost threshold is defined as the value of $b/c$ at which $\mathcal{C}^Y_{Xn}(\delta, \varepsilon) = 0.5$, denoted $(b/c)^Y_{Xn}(\delta, \varepsilon)$. In the scenarios considered here, cooperation exhibits the following pattern (see Supplementary Information Section 2.3). If the threshold exceeds $1$, cooperation emerges when $b/c > (b/c)^Y_{Xn}(\delta, \varepsilon)$. If the threshold is below $1$, $\mathcal{C}^Y_{Xn}(\delta, \varepsilon) > 0.5$ occurs when $b/c < (b/c)^Y_{Xn}(\delta, \varepsilon)$, corresponding either to insufficient cooperation ($0 < b/c < 1$) or to spiteful behavior ($b/c < 0$, in which individuals pay a cost to harm others \cite{hamilton1970selfish}).

To investigate how repeated interactions influence cooperative outcomes, we consider the idealized case of infinitely repeated, error-free games ($\delta = 1$, $\varepsilon = 0$), focusing on two deterministic strategy sets: reactive-$1$ and memory-$1$. The critical thresholds are (confirmed by simulations, Fig.~\ref{simul_det_inf_no}):
\begin{equation}
    \begin{aligned}
        &(b/c)_R^\ast(1,0) = \frac{5\tau + 1}{\tau + 5}, \\
        &(b/c)_M^\ast(1,0) = \frac{529\tau + 239}{239\tau + 529}.
    \end{aligned}
\end{equation}
Above, $\tau$ is the network coefficient quantifying the effect of population structure, with $|\tau| > 1$. For one-shot games, the threshold reduces to $(b/c)^\ast = \tau$ \cite{allen2017evolutionary}.

\begin{figure}[!t]
	\centering
	\includegraphics[width=\textwidth]{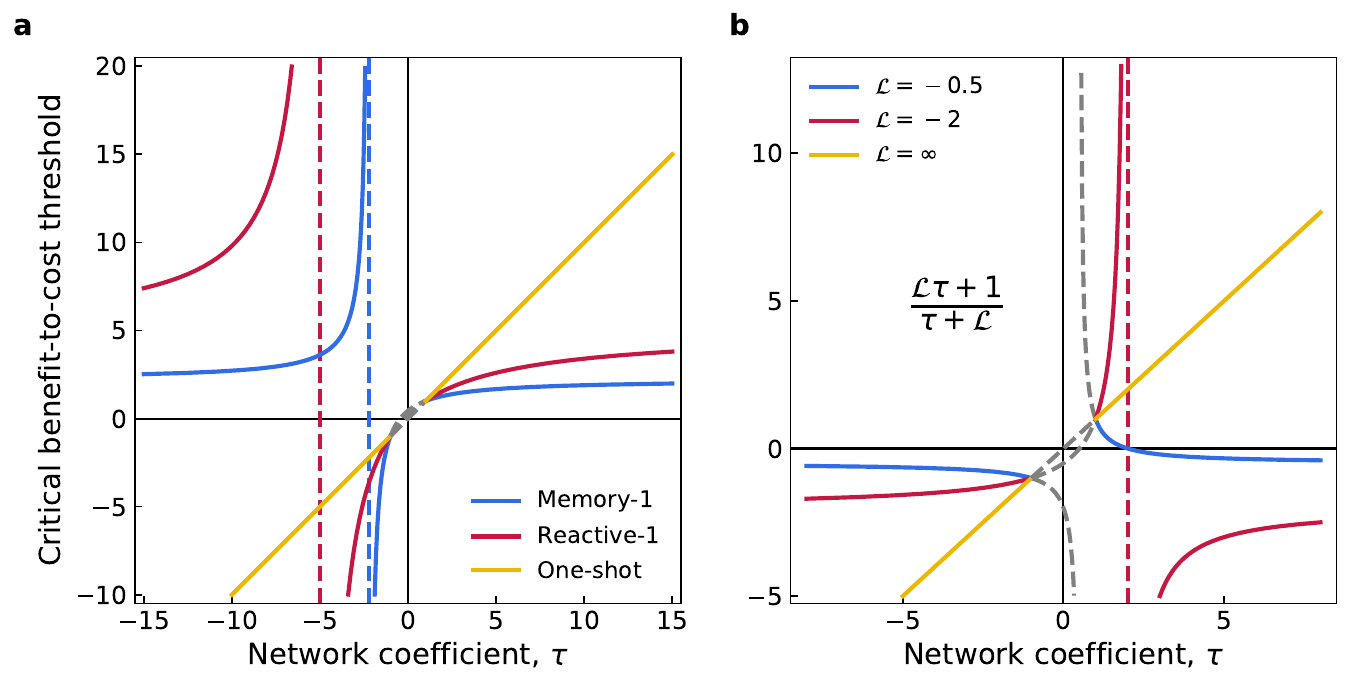}
	\caption{\textbf{A full spectrum for the evolution of cooperation under iterated and one-shot games on networks.} \textbf{a}, Thresholds in infinitely repeated, error-free games relative to one-shot thresholds, plotted against the network coefficient $\tau$. \textbf{b}, Thresholds $(\mathcal{L}\tau+1)/(\tau+\mathcal{L})$ plotted against $\tau$ for selected $\mathcal{L}$. Cases with $\mathcal{L}>1$ correspond to the red ($\mathcal{L}=5$) and blue ($\mathcal{L} \approx 2.2$) lines in panel~\textbf{a}. Gray dashed lines mark the unattainable region $|\tau| \le 1$.}
	\label{spectrum}
\end{figure}

Figure~\ref{spectrum}\textbf{a} compares thresholds for one-shot and iterated games across network structures, where each value of $\tau$ corresponds to a distinct structural class. When $\tau > 1$, repeated interactions lower the threshold, making cooperation more attainable. When $\tau < -1$, repeated interactions can either promote cooperation or suppress spite. Notably, repeated interactions shift the threshold from negative to positive when $\tau < -5$ (reactive-$1$) or $\tau < -2.2$ (memory-$1$), thereby enabling cooperation in networks that would otherwise favor spite. In this context, networks with more negative values of $\tau$, which are more inclined toward spite in one-shot settings, display thresholds that are more favorable to cooperation under repeated interactions. For moderate negative $\tau$, repeated interactions help mitigate spiteful behavior, though they may not suffice to foster cooperation.

\subsection*{A unifying indicator for cooperation in repeated interactions} 
To disentangle the effects of repeated-interaction strategies from population structure, we introduce a unifying indicator $\mathcal{L}_{Xn}^Y(\delta, \varepsilon)$. It applies to any strategy class for which cooperation rates and payoffs can be computed, and thus extends beyond the specific classes examined here.
Using this indicator, we can rewrite the threshold as (see Supplementary Information, Sections 2.1–2.2):
\begin{equation}
    (b/c)_{Xn}^Y(\delta,\varepsilon) = \frac{\mathcal{L}_{Xn}^Y(\delta,\varepsilon) \times \tau + 1}{\tau + \mathcal{L}_{Xn}^Y(\delta,\varepsilon)}.
\end{equation}

Both the sign and magnitude of $\mathcal{L}$ determine how repeated interactions affect cooperation. Across all scenarios considered, $\mathcal{L}$ falls into one of three categories: (i) when $\mathcal{L} = \infty$, repeated interactions have no influence on cooperation, the threshold coincides with the one-shot value $\tau$; (ii) when $\mathcal{L} > 1$, repeated interactions promote cooperation, with smaller values of $\mathcal{L}$ corresponding to greater cooperative benefit; (iii) when $\mathcal{L} < 0$, repeated interactions undermine cooperation.

These regimes are illustrated in Fig.~\ref{spectrum}. In panel~\textbf{a}, curves correspond to different indicator values with $\mathcal{L}_1 > \mathcal{L}_2 > 1$. When $\tau > 1$ or $\tau < -\mathcal{L}_1$, both scenarios yield positive thresholds, but the case with $\mathcal{L}_2$ produces a lower threshold, indicating that cooperation is more easily achieved. When $-\mathcal{L}_1 < \tau < -\mathcal{L}_2$, only the smaller $\mathcal{L}_2$ still supports cooperation (i.e., yields a positive threshold), while $\mathcal{L}_1$ does not. Panel~\textbf{b} depicts the case with $\mathcal{L} < 0$: if $\mathcal{L} < -1$, repeated interactions inhibit cooperation for $\tau > 1$ and promote spite when $\tau < -1$. If $-1 < \mathcal{L} < 0$, repeated interactions allow only insufficient cooperation in network structures with $1 < \tau < -1/\mathcal{L}$.

\subsection*{Robustness of cooperation to error and continuation probability}
\begin{figure}[]
	\centering
	\includegraphics[width=\textwidth]{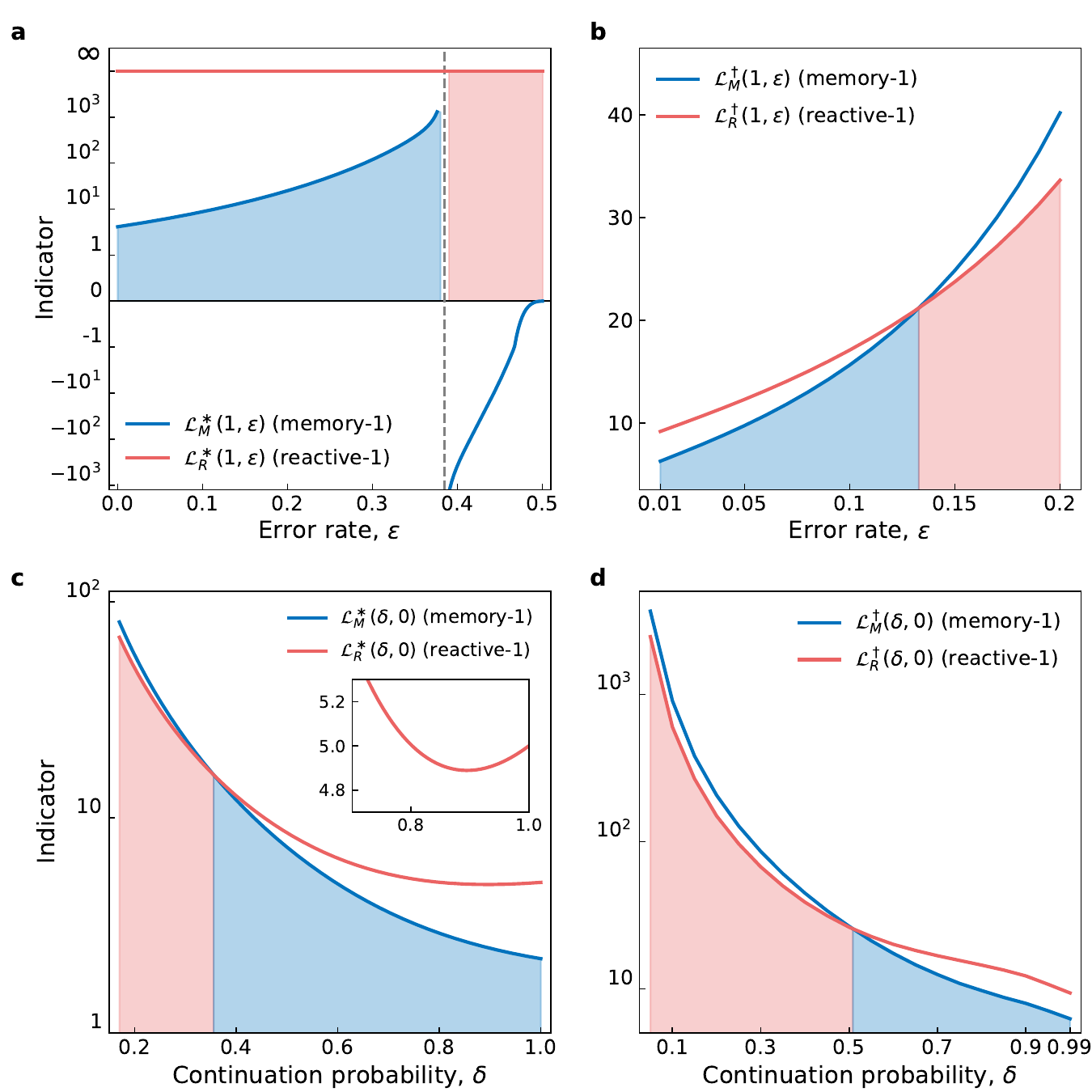}
	\caption{\textbf{The cooperative advantage of repeated interactions is robust to error rates and continuation probabilities.} \textbf{a}, \textbf{b}, Indicator values for memory-$1$ (blue) and reactive-$1$ (red) strategies plotted against the error rate $\varepsilon$ under deterministic (\textbf{a}) and stochastic (\textbf{b}) settings. Blue (red) shading marks the range of $\varepsilon$ values where memory-$1$ (reactive-$1$) strategies more effectively promote cooperation. \textbf{c}, \textbf{d}, Indicator values for memory-$1$ (blue) and reactive-$1$ (red) strategies plotted against the continuation probability $\delta$ under deterministic (\textbf{c}) and stochastic (\textbf{d}) settings. Blue (red) shading marks the range of $\delta$ values where memory-$1$ (reactive-$1$) strategies more effectively promote cooperation.} 
	\label{error+finite}
\end{figure}

We begin by examining the impact of implementation errors. In the absence of errors, the indicator for deterministic reactive-$1$ strategies equals $5$, meaning that repeated interactions are strongly conducive to cooperation. As soon as any error is introduced, however, this value diverges to infinity, indicating that repeated interactions cease to provide a cooperative advantage. (Fig.~\ref{error+finite}\textbf{a}). This sharp transition demonstrates the detrimental effect of errors on cooperation. Results for other strategy classes further support this conclusion. For deterministic memory-$1$ strategies, the indicator is positive at low error rates but turns negative as errors become more frequent. For both stochastic memory-$1$ and reactive-$1$ strategies (Fig.~\ref{error+finite}\textbf{b}), the indicator increases steadily with the error rate. Together, these patterns indicate that high error rates progressively compromise the benefits of repeated interactions. Such impairment renders cooperation fragile in noisy environments and underscores the importance of precise strategy execution. This pattern persists for strategies with longer memory (SI~Fig.~$5$).

We next investigate how continuation probability shapes cooperative dynamics. For deterministic memory-$1$ strategies, the indicator declines monotonically as the continuation probability increases (Fig.~\ref{error+finite}\textbf{c}), consistent with the intuition that longer relationships strengthen reciprocity (since smaller $\mathcal{L}$ values correspond to a stronger cooperative advantage). A similar pattern holds for stochastic reactive-$1$ and memory-$1$ strategies (Fig.~\ref{error+finite}\textbf{d}). The only exception occurs for deterministic reactive-$1$ strategies, whose indicator exhibits a non-monotonic pattern with a minimum near $\delta \approx 0.9$. This type of non-monotonicity is not observed for any other strategy class. In all remaining cases, the indicator $\mathcal{L}$ decreases steadily as continuation probability increases (SI~Fig.~$7$), reinforcing the conclusion that higher continuation probabilities reliably favor cooperation.

\subsection*{Effects of strategy features on cooperation}
\begin{figure}[!t]
	\centering
	\includegraphics[width=.5\textwidth]{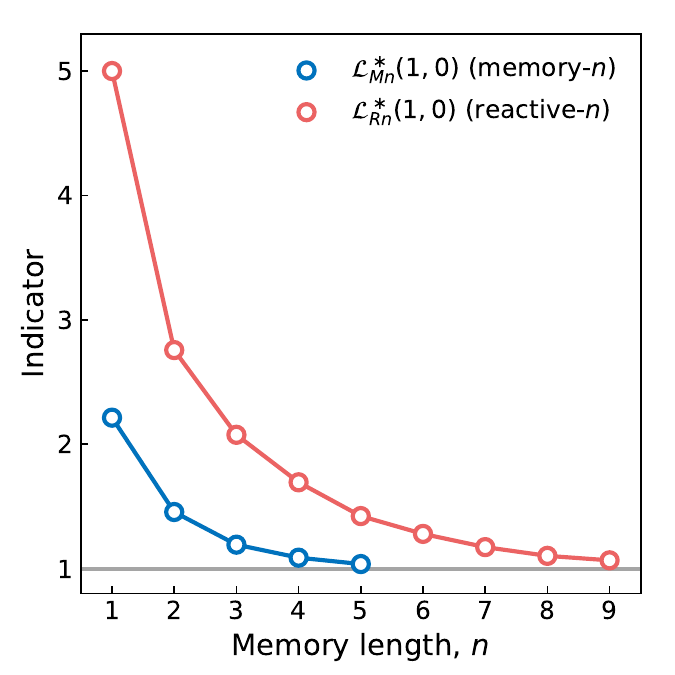}
	\caption{\textbf{Cooperative advantage of extended memory under infinite repetition.} Indicator values for memory-$n$ (blue) and reactive-$n$ (red) strategies plotted against the memory length $n$ in infinitely repeated, error-free games. In both cases, indicator values decrease with $n$, suggesting that longer memory supports cooperation. Memory-$n$ strategies consistently yield lower indicator values than reactive-$n$ strategies.} 
	\label{memory_length}
\end{figure}

\begin{figure}[!t]
	\centering
	\includegraphics[width=\textwidth]{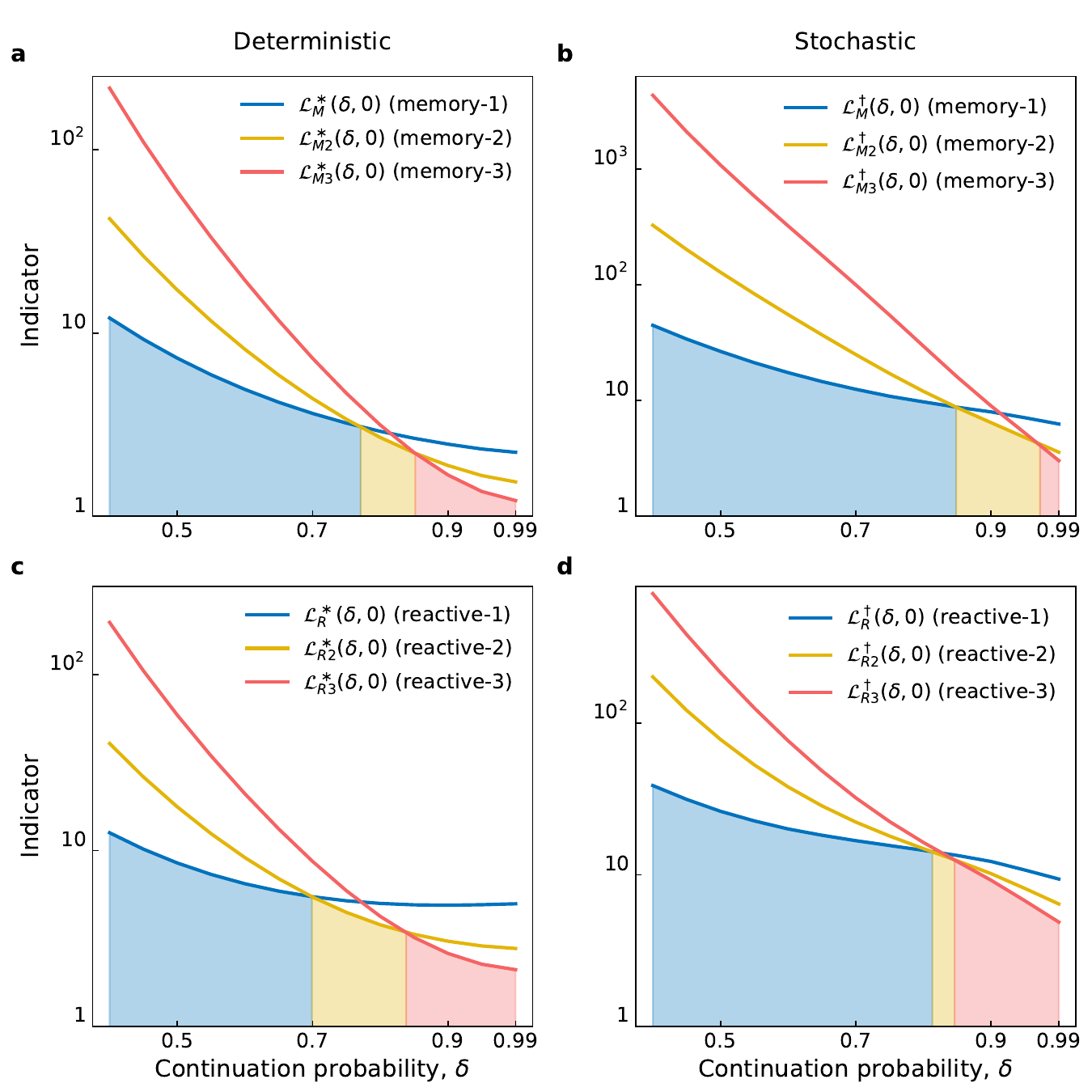}
	\caption{\textbf{Conditional advantage of memory across continuation probabilities.} 
    \textbf{a}-\textbf{d}, Indicator values plotted against the continuation probability $\delta$ for strategies with memory lengths $n = 1$ (blue), $n = 2$ (yellow), and $n = 3$ (red). Columns show deterministic (left) and stochastic (right) strategies, while rows show memory-$n$ (top) and reactive-$n$ (bottom) strategies. Shaded regions indicate the memory lengths $n$ that yield the lowest indicator values (i.e., most conducive to cooperation) across values of $\delta$. As continuation probability increases, strategies with longer memory become more effective.}
	\label{delta_memory_length}
\end{figure}

Memory length plays a central role in shaping the cooperative potential of strategies. In infinitely repeated games, indicator values decline as memory length increases (Fig.~\ref{memory_length}), indicating that longer memory promotes cooperation (see SI~Fig.~$8$ for the error-prone scenario). However, the advantage of extended memory depends on the likelihood of further interaction (Fig.~\ref{delta_memory_length}): when continuation probability is low, interactions are typically brief, providing insufficient history for strategies with longer memory to use effectively; as repeated interactions become more likely, extended memory enables strategies to condition behavior on more detailed past information, thereby enhancing cooperation.

\begin{figure}[!t]
	\centering
	\includegraphics[width=\textwidth]{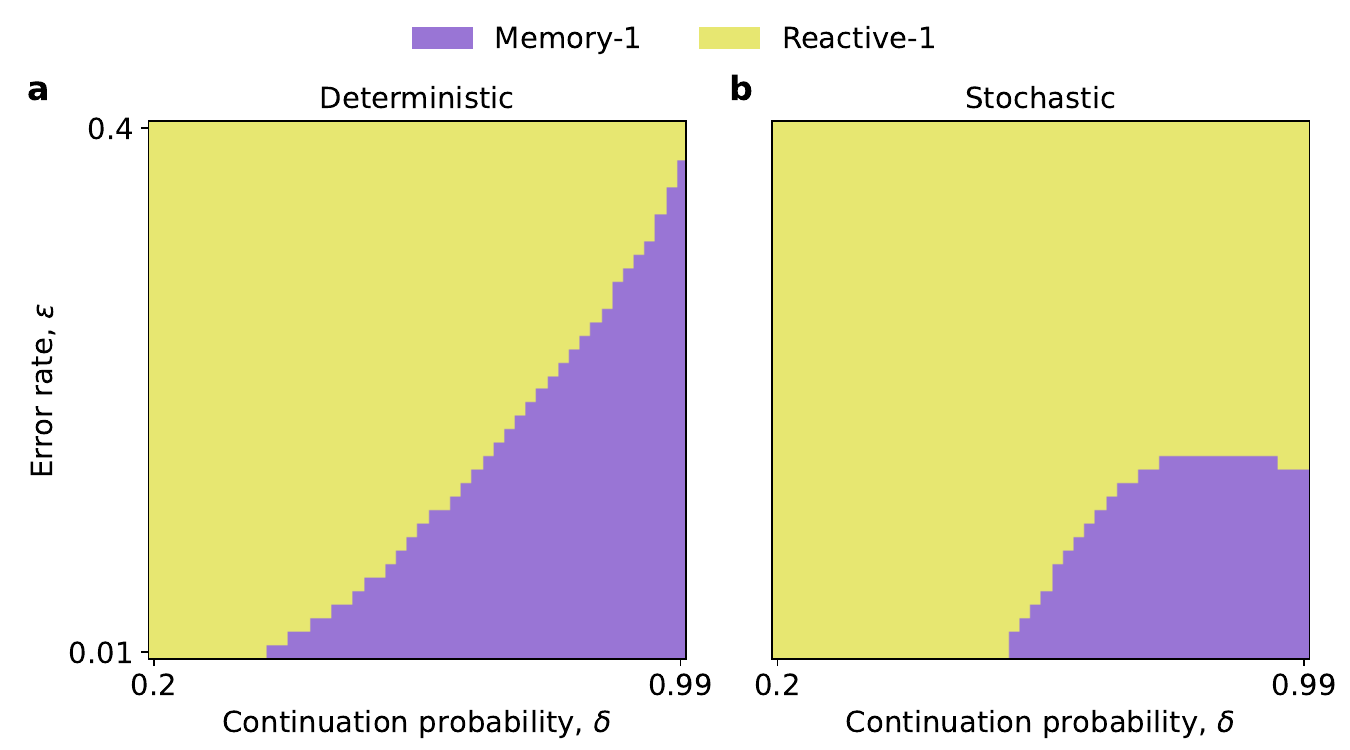}
	\caption{\textbf{Trade-offs between memory-$1$ and reactive-$1$ strategies in promoting cooperation.} \textbf{a}, \textbf{b}, Comparison of memory-$1$ and reactive-$1$ strategies in promoting cooperation for deterministic (\textbf{a}) and stochastic (\textbf{b}) settings. Purple regions indicate where memory-$1$ strategies are more effective; yellow regions indicate an advantage for reactive-$1$ strategies. Memory-$1$ strategies promote cooperation more effectively than reactive-$1$ strategies when the continuation probability is high and the error rate is low, whereas reactive-$1$ strategies are more effective in other conditions.}
	\label{delta_error_memory_reactive_one}
\end{figure}

Informational input further modulates performance. Reactive-$n$ strategies rely solely on the opponent’s past actions, whereas memory-$n$ strategies incorporate the histories of both players. This broader informational basis enables memory-$n$ strategies to support cooperation more effectively under favorable conditions. For instance, memory-$1$ strategies outperform reactive-$1$ strategies at low error rates (Fig.~\ref{error+finite}\textbf{a},\textbf{b}). However, as error rates increase, reactive-$1$ strategies become more advantageous, maintaining cooperation more effectively despite relying on simpler information. A similar contrast appears with continuation probability: when the expected interaction is short, reactive-$1$ strategies perform better, whereas memory-$1$ strategies are more effective when long-term relationships are expected (Figs.~\ref{error+finite}\textbf{c},\textbf{d} and \ref{memory_length}). These trade-offs are summarized in Fig.~\ref{delta_error_memory_reactive_one}, which compares the relative performance of memory-$1$ and reactive-$1$ strategies across varying error rates and continuation probabilities. This pattern is consistent for strategies with longer memory as well (SI~Figs.~$5$ and~$7$).

Finally, decision rules critically influence cooperative outcomes. Deterministic strategies, which respond consistently to a given history, generally promote cooperation more effectively than stochastic strategies, which introduce randomness into decision-making. As shown in SI~Figs.~$9$ and~$10$, this advantage holds across a wide range of error rates and continuation probabilities. The predictability of deterministic strategies improves coordination and fosters stable cooperation, whereas the variability of stochastic ones introduces uncertainty that can undermine cooperative behavior.

\section*{Discussion}
Repeated interactions shape cooperative behavior by allowing individuals to adjust future actions based on past experiences. Decades of research have identified influential strategies such as Tit-for-Tat, Generous Tit-for-Tat, and Win-Stay Lose-Shift. Yet these well-known strategies occupy only a small region of the vast strategy space, and success in isolated pairwise contests does not necessarily translate into evolutionary success at the population level. Moreover, most prior studies focus on well-mixed populations, where individuals interact randomly. This focus overlooks the richer dynamics that emerge in structured populations. Here, we establish a general framework for repeated-interaction strategies that accommodates arbitrary memory lengths, diverse informational inputs, and both deterministic and stochastic decision rules. This framework enables a systematic analysis of how repeated interactions influence the emergence and stability of cooperation in structured populations.

Our theoretical analysis identifies a critical benefit-to-cost threshold above which cooperation is favored by natural selection. This threshold depends on two quantities: the network coefficient $\tau$, which captures the influence of population structure; and the indicator $\mathcal{L}$, which reflects the combined effects of the strategy set, continuation probability, and error rate. Comparing this threshold with that of the one-shot game reveals that repeated interactions substantially reduce the barrier for cooperation, particularly in networks where cooperation is impossible under one-shot interactions.

These findings extend prior work in several directions. First, although previous studies have shown that network structure can promote cooperation in one-shot games \cite{allen2017evolutionary}, this effect vanishes in networks with weak structural reinforcement ($\tau < -1$). By contrast, our results show that repeated interactions can foster cooperation in a much broader class of networks, including those with $\tau < -\mathcal{L}$ and $\tau > 1$. This offers a potential explanation for the prevalence of cooperation in real systems, where it often emerges even in relatively dense networks, which are typically associated with $\tau < -1$, defying predictions from one-shot frameworks. Moreover, for any network with $\tau > 1$, the indicator $\mathcal{L}$ defines a universal benefit-to-cost threshold above which cooperation is favored, underscoring the broad capacity of repeated interactions to promote cooperation across diverse structures. Second, population structure itself exerts a decisive influence. In well-mixed populations, as assumed in earlier studies, each pair of individuals interacts with equal probability, which corresponds to a complete graph with network coefficient $\tau = -(N - 1)$. We show that networks with either $\tau < -(N-1)$ or $\tau > 1$ are more conducive to cooperation under repeated interactions than the complete graph. Third, unlike much of the existing literature, which has relied primarily on simulations \cite{tkadlec2023mutation, donahue2020evolving, li2022evolution, baek2016comparing}, we develop a general theoretical framework from which explicit analytical conditions can be derived. The indicator $\mathcal{L}$ quantifies the impact of repeated-interaction strategies independently of the underlying network, enabling standardized comparisons across strategy families. In most cases, $\mathcal{L} > 1$, with smaller values indicating greater capacity to sustain cooperation. As such, $\mathcal{L}$ serves as a concise and interpretable metric, allowing for direct comparison across diverse strategy classes.

We apply the indicator $\mathcal{L}$ to assess how the key features of repeated-interaction strategies influence cooperative dynamics. Among these features, memory length emerges as a primary determinant. Longer-memory strategies can sustain cooperation even in noisy environments by detecting patterns and enforcing reciprocity. However, their advantage diminishes when the expected duration of interactions is short. Informational input is likewise crucial. Memory-$n$ strategies, which consider both players’ past actions, typically outperform reactive-$n$ strategies, which rely solely on the opponent’s past moves, under low error rates and high continuation probabilities. In contrast, reactive-$n$ strategies show greater robustness in noisy or short-term settings. Decision rules constitute another critical factor. Deterministic strategies produce consistent responses that enhance predictability and coordination, whereas stochastic strategies introduce variability that can undermine reciprocity and weaken cooperative stability.

These results refine and, in some respects, challenge established conclusions. Earlier work suggested that longer memory offers no advantage in direct competition with short-memory strategies \cite{press2012iterated}, leading many studies to focus exclusively on memory-$1$ strategies \cite{donahue2020evolving, hilbe2018evolution, tkadlec2023mutation}. Our findings show instead that in evolving populations, longer memory can promote cooperation under favorable conditions, particularly when continuation probability is high. This contradicts the traditional view that extended memory offers no advantage, revealing that its benefits arise not from head-to-head contests but from the cumulative effects of overlapping interactions across the population. Thus, the consequences of repeated interactions extend beyond dyadic outcomes to the broader evolutionary landscape: cooperation depends less on individual payoffs and more on collective patterns shaped by population-level strategy interactions. This perspective is further supported by the observed relationship between strategy characteristics and robustness. Tit-for-Tat, a deterministic reactive-$1$ strategy, is fragile under noise, inspiring refinements designed to enhance resilience. One such refinement is Win-Stay Lose-Shift, a memory-$1$ strategy adapted for noisy settings \cite{nowak1993strategy}, often taken to suggest the superiority of memory-$1$ under noise. Yet we find that in evolving populations, reactive-$n$ strategies can be more robust than memory-$n$ in such environments. Another refinement is Generous Tit-for-Tat \cite{nowak1992tit}, a stochastic variant designed to reduce brittleness, frequently cited as evidence for the value of stochasticity. Nevertheless, our results indicate that deterministic strategies tend to sustain cooperation more effectively than stochastic ones. Together, these findings reveal the limits of pairwise-contest analysis and underscore the importance of considering structured, evolving populations.

This work suggests several avenues for future research. While our analysis has centered on classical strategy classes, it is far from exhaustive. Future work could extend the framework to encompass more complex strategies, such as those capable of tracking complete interaction histories \cite{li2022evolution} or those derived from algorithmic and learning-based approaches \cite{harper2017reinforcement}, which may offer deeper insights into the mechanisms and evolutionary logic of repeated interactions. In addition, although our model considers only pairwise interactions, real-world social systems often involve strategic interactions among multiple individuals, including collective decision-making and group coordination. A hypergraph-based framework could provide a more faithful representation of such settings \cite{sheng2024strategy, alvarez2021evolutionary, boccaletti2023structure, civilini2024explosive, gao2025evolutionary}. These avenues offer promising directions for deepening our understanding of the mechanisms that sustain cooperation in complex social environments.

\section*{Methods}
\subsection*{Interaction outcomes for strategies with one-step memory}
\subsubsection*{Limiting distribution}
We begin by outlining how to compute the long-term outcomes when two individuals interact using memory-$1$ strategies. A memory-$1$ strategy is represented as $\bar{\mathbf{p}} = (p_0; \mathbf{p})$, where $p_0$ is the probability of cooperating in the first round, and $\mathbf{p} = (p_{\mathrm{CC}}, p_{\mathrm{CD}}, p_{\mathrm{DC}}, p_{\mathrm{DD}})$ specifies the probability of cooperating based on the previous round’s outcome. The opponent’s strategy, $\bar{\mathbf{q}} = (q_0; \mathbf{q})$, is defined analogously.

In infinitely repeated games ($\delta = 1$), the resulting dynamics depend on whether implementation errors occur. When errors are absent ($\varepsilon = 0$), we focus on deterministic strategies. In this case, the Markov process may not have a unique stationary distribution but eventually enters a deterministic cycle. This cycle yields a limiting distribution, $\mathbf{v} = (v_{\mathrm{CC}}, v_{\mathrm{CD}}, v_{\mathrm{DC}}, v_{\mathrm{DD}})$, which can be computed algorithmically.

When errors are present ($0 < \varepsilon \le 0.5$), the process is ergodic: the limiting distribution is unique and independent of the initial state. It is given by the left eigenvector associated with the eigenvalue $1$ of the transition matrix:
\begin{equation}
	\mathbf{M}_{\tilde{\mathbf{p}}\tilde{\mathbf{q}}}=\left(\begin{array}{llll}
		\tilde{p}_{\mathrm{CC}} \tilde{q}_{\mathrm{CC}} & \tilde{p}_{\mathrm{CC}}\left(1-\tilde{q}_{\mathrm{CC}}\right) & \left(1-\tilde{p}_{\mathrm{CC}}\right) \tilde{q}_{\mathrm{CC}} & \left(1-\tilde{p}_{\mathrm{CC}}\right)\left(1-\tilde{q}_{\mathrm{CC}}\right) \\
		\tilde{p}_{\mathrm{CD}} \tilde{q}_{\mathrm{DC}} & \tilde{p}_{\mathrm{CD}}\left(1-\tilde{q}_{\mathrm{DC}}\right) & \left(1-\tilde{p}_{\mathrm{CD}}\right) \tilde{q}_{\mathrm{DC}} & \left(1-\tilde{p}_{\mathrm{CD}}\right)\left(1-\tilde{q}_{\mathrm{DC}}\right) \\
		\tilde{p}_{\mathrm{DC}} \tilde{q}_{\mathrm{CD}} & \tilde{p}_{\mathrm{DC}}\left(1-\tilde{q}_{\mathrm{CD}}\right) & \left(1-\tilde{p}_{\mathrm{DC}}\right) \tilde{q}_{\mathrm{CD}} & \left(1-\tilde{p}_{\mathrm{DC}}\right)\left(1-\tilde{q}_{\mathrm{CD}}\right) \\
		\tilde{p}_{\mathrm{DD}} \tilde{q}_{\mathrm{DD}} & \tilde{p}_{\mathrm{DD}}\left(1-\tilde{q}_{\mathrm{DD}}\right) & \left(1-\tilde{p}_{\mathrm{DD}}\right) \tilde{q}_{\mathrm{DD}} & \left(1-\tilde{p}_{\mathrm{DD}}\right)\left(1-\tilde{q}_{\mathrm{DD}}\right)
	\end{array}\right),
\end{equation}
where $\tilde{\mathbf{p}}=(1-2\varepsilon)\mathbf{p}+\varepsilon$ and $\tilde{\mathbf{q}}=(1-2\varepsilon)\mathbf{q}+\varepsilon$ take into account implementation errors.

When $\delta < 1$, the game has a finite expected length. In this case, the limiting distribution depends both on the transition matrix and on the initial state and is given by:
\begin{equation}
	\mathbf{v}=(1-\delta) \mathbf{v}_0(\mathbf{I}-\delta \mathbf{M}_{\tilde{\mathbf{p}}\tilde{\mathbf{q}}})^{-1},
\end{equation}
with the initial distribution over states defined as:
\begin{equation}
    \mathbf{v}_0=\left(p_{0}q_{0}, p_{0}\left(1-q_{0}\right), \left(1-p_{0}\right)q_{0}, (1-p_{0})(1-q_{0})\right).
\end{equation}

\subsubsection*{Cooperation rate and payoff}
Given the limiting distribution $\mathbf{v}$, the expected payoff of strategy $\bar{\mathbf{p}}$ against $\bar{\mathbf{q}}$ is:
\begin{equation}
	\pi(\bar{\mathbf{p}},\bar{\mathbf{q}})=(b-c) v_{\mathrm{CC}} + (-c) v_{\mathrm{CD}} + b v_{\mathrm{DC}},
\end{equation}
where $b$ and $c$ are the benefit and cost of cooperation, respectively. The cooperation rate is defined as:
\begin{equation}
	\mathcal{C}(\bar{\mathbf{p}},\bar{\mathbf{q}})=v_{\mathrm{CC}} + v_{\mathrm{CD}}.
\end{equation}

\subsection*{Extension to strategies with longer memory}
For strategies with memory length $n$, players choose actions based on the outcomes of the last $n$ rounds \cite{glynatsi2024conditional}. A history is denoted by $\mathbf{h} = (a_1\tilde{a}_1, \ldots, a_n\tilde{a}_n)$, where $a_t$ and $\tilde{a}_t$ are the actions of the focal player and opponent $t$ rounds ago, respectively.

The sets of deterministic and stochastic memory-$n$ strategies are defined as:
\begin{equation}
	\begin{aligned}
		&\mathcal{S}^\ast_{Mn}=\left\{(\mathcal{A}^\ast_n,\mathbf{p}) \Big| \mathbf{p}=\{p_\mathbf{h}\}_{\mathbf{h}} \in \{0,1\}^{2^{2^n}}\right\},\\
		&\mathcal{S}^\dagger_{Mn}=\left\{(\mathcal{A}^\dagger_n,\mathbf{p}) \Big| \mathbf{p}=\{p_\mathbf{h}\}_{\mathbf{h}} \in [0,1]^{2^{2^n}}\right\},
	\end{aligned}
\end{equation}
where $\mathcal{A}_n^\ast$ and $\mathcal{A}_n^\dagger$ specify the distributions over initial moves (deterministic and stochastic, respectively), and $p_\mathbf{h}$ denotes the probability of cooperating conditional on history $\mathbf{h}$. For reactive-$n$ strategies, the probability of cooperation depends only on the opponent’s previous $n$ actions.

The procedures for calculating limiting distributions, cooperation rates, and payoffs extend directly to these longer-memory strategies.

\subsection*{Cooperation rate for the population}
At the population level, we define the cooperation rate in the limit of rare exploration ($\mu \rightarrow 0$). In this regime, the population is typically monomorphic, with occasional exploration introducing a new strategy that either fixates or goes extinct before the next exploration event. As a result, the long-run cooperation rate can be approximated by averaging the cooperation levels of strategies that successfully fixate. This rate depends on the strategy set $\mathcal{S}^Y_{Xn}$, continuation probability $\delta$, and error rate $\varepsilon$.

For the deterministic strategy set $\mathcal{S}^\ast_{Xn}$, the cooperation rate is:
\begin{equation}
	\mathcal{C}^\ast_{Xn}(\delta,\varepsilon) = \sum_{\mathbf{p} \in \mathcal{S}^\ast_{Xn}}u_{\mathbf{p}}(\delta,\varepsilon) \times \mathcal{C}_{\mathbf{p}}(\delta,\varepsilon),
\end{equation}
where $\mathcal{C}_{\mathbf{p}}(\delta, \varepsilon)$ is the cooperation rate of strategy $\mathbf{p}$ against itself, and $u_{\mathbf{p}}(\delta, \varepsilon)$ is its stationary frequency. 

For the stochastic strategy set $\mathcal{S}^\dagger_{Xn}$, the cooperation rate is:
\begin{equation}
	\mathcal{C}^\dagger_{Xn}(\delta,\varepsilon) = \int_{\mathbf{p} \in \mathcal{S}^\dagger_{Xn}}\text{f}_{\mathbf{p}}(\delta,\varepsilon) \times \mathcal{C}_{\mathbf{p}}(\delta,\varepsilon) \text{d}\mathbf{p},
\end{equation}
where $\text{f}_{\mathbf{p}}(\delta,\varepsilon)$ is the density of strategy $\mathbf{p}$ under the stationary distribution.

\bibliographystyle{unsrt}
\bibliography{reference}

\end{document}